# Dosimetric impact of physician style variations in contouring CTV for postoperative prostate cancer: A deep learning–based simulation study

Anjali Balagopal, Dan Nguyen, Maryam Mashayekhi, Howard Morgan, Aurelie Garant, Neil Desai, Raquibul Hannan, Mu-Han Lin, Steve Jiang

Medical Artificial Intelligence and Automation (MAIA) Laboratory and Department of Radiation Oncology, UT Southwestern Medical Center

**Corresponding Author**:

Steve.Jiang@UTSouthwestern.edu,+1-214-645-8542, Department of Radiation Oncology, University of Texas Southwestern Medical Center, 2280 Inwood Road, Dallas, TX 75390-9303

**Total word count (excluding Abstracts, figures, tables, and references**): 3543

**Total table and figure count:** 8

# ABSTRACT

Inter-observer variation is a significant problem in clinical target volume (CTV) segmentation in postoperative settings, where there is no gross tumor present. In this scenario, the CTV is not an anatomically established structure, but one determined by the physician based on the clinical guideline used, the preferred tradeoff between tumor control and toxicity, their experience and training background, and other factors. This results in high inter-observer variability between physicians. This variability has been considered an issue, but the absence of multiple physician CTV contours for each patient and the significant amount of time required for dose planning based on each physician CTV have made it impractical to study its dosimetric consequences. In this study, we analyze the impact that variations in physician style have on dose to organs-at-risk (OAR) by simulating the clinical workflow via deep learning. For a given patient previously treated by one physician, we use deep learning–based tools to simulate how other physicians would contour the CTV and how the corresponding dose distributions would look for other physicians. To simulate multiple physician styles, we use a previously developed in-house CTV segmentation model that can produce physician style–aware segmentations. The corresponding dose distribution is predicted using another in-house deep learning tool, which, **can predict dose within 3% of the prescription dose, on average, on the test data.** For every test patient, four different physician style CTVs are considered, and four different dose distributions are analyzed. OAR dose metrics are compared, showing that even though physician style variations result in organs getting different doses, all the important dose metrics except Maximum Dose point are within the clinically acceptable limit.

# 1. INTRODUCTION

Optimal radiation treatment entails uniform full-dose coverage of the radiation target with a sharp dose fall-off around it. This necessitates precise segmentation of both the radiation target and the nearby organs that are at risk for radiation damage (organs-at-risk, OARs). In a typical radiotherapy setting, OARs are segmented together with the gross tumor volume (GTV), which is the tumor that is visible in images. Using their knowledge of the disease, physicians then expand the GTV to create the clinical target volume (CTV), which also includes the microscopic extensions not visible in images. In the case of postoperative radiotherapy, however, the visible tumor has been surgically removed, so the CTV is only a virtual volume encompassing an area that may contain microscopic tumor cells, not an expansion of a macroscopic or visible tumor volume. The treating physician determines the optimal CTV based not only on patient characteristics but also on potential dose concerns for OARs. Physicians decide on the CTV contour based on their experience regarding toxicity, especially since many surrounding organs form part of the CTV in postoperative cases. In situations where CTV is contoured based on potential dose concerns, the CTV segmentation is impacted by high observer variability [1-6]. A previous study conducted in-house showed that distinguishable CTV contouring style exists between physicians, but these style variations were not found to have an impact on patient outcome [7].

Physicians expand the CTV to a planning tumor volume (PTV), which encloses the CTV with margins to account for possible uncertainties in beam alignment, patient positioning and organ motion. Ideally, the CTV–PTV margin should be determined solely by the magnitudes of the uncertainties involved. In practice, physicians usually also consider doses to nearby healthy tissues when deciding on the size of the margin. When dose planning is performed with this PTV that does not use recommended margins (that are based on setup errors), the CTV that is being treated would end up being different from what was originally contoured. This introduces additional variations. The impact of the style variations in the CTV, along with the added variations due to the non-ideal margins used for PTV, needs to be further studied from a dosimetric point of view.

Treatment planning is a time-consuming process consisting of manual segmentation of required tumor and organ contours and multiple rounds of parameter tuning and optimization in a trial-and-error manner, which has so far prevented rigorous dosimetric studies of variations in the CTV. Automating the treatment planning process with the help of available historical data can accelerate treatment planning. Multiple studies have used deep learning (DL) to successfully segment CTVs and organs [8-13]. Many radiotherapy dose prediction studies have also been reported and can be broadly separated into two categories. The first category deals with dose-volume histogram (DVH) prediction based on mathematical frameworks [14, 15]. These studies focus only on DVH indices as model inputs and outputs without three-dimensional dose volume information. They provide information for potentially optimal plans that spare OARs as much as possible. The second category deals with three-dimensional dose distribution prediction using deep learning networks [16-23]. These studies have shown relatively high prediction precision according to the quantitative index comparisons, and the predicted results, which are three-dimensional dose volumes, could provide probabilities for all quantitative and qualitative evaluations. Several studies have addressed dose distribution prediction using deep learning for prostate dose prediction such as prostate SBRT [24], prostate IMRT[25] and prostate VMAT[26]. These models use UNet based architectures with MSE loss and works on preoperative patient data.

In this study, we analyze the impact that variations in physician CTV segmentation style have on dose to organs-at-risk (OARs). For post-operative prostate patients, the dose to OAR is both physician and patient dependent. Disentangling the physician vs patient dependent features is difficult in this scenario. For every patient undergoing radiation therapy treatment, a single physician segments the CTV and finalizes the dose planning. A set of patients treated by one physician cannot be directly compared to a different set of patients treated by another physician, since the patient anatomy, individual characteristics as well as comorbidities could be the cause for any difference detected. It would be beneficial to compare same set of patients across physicians. To evaluate the dosimetric impact of physician styles, for the same patient, CTVs drawn by the rest of the physicians and the corresponding dose plans are required. Contouring multiple CTVs per patient and further performing dose planning for each of these CTVs is time consuming and cumbersome since treatment planning for one patient takes at the least half a day. To analyze the dosimetric impact that variations in CTV segmentation styles across physicians, dose planning will have to be performed individually for each of the physicians' CTVs. In this study we propose the use of DL models for mimicking the clinical workflow and enabling the generation of multiple dose plans based on multiple physician CTVs for each patient.

A previously developed in-house physician style CTV segmentation model [7] that can predict CTVs for each patient in 4 different physicians' styles is used for generating physician specific CTVs. This model can generate CTVs in four different contouring styles using a multi-head attention network and perceptual loss. For each patient, we already have the CTV contour manually segmented by one physician that was used to create the ground truth dose plan and treat the patient, so the segmentation model is used to generate CTVs in the other three physicians' styles. Given that physicians do not always use the same margins for CTV–PTV expansion, we use individual DL based PTV expansion networks for each physician to mimic how each physician would expand the CTV to the PTV with the help of OAR contours. This step is necessary to mimic the current clinical workflow and emulate how each physician would expand the CTV to PTV. We show how the PTV expansion model significantly improves the CTV–PTV expansion accuracy over a nominal expansion using recommended margins. Additionally, we developed a dose prediction model for predicting postoperative prostate doses based on the organ masks and the generated PTV masks. For a set of test patients, the DVH and dose metrics are calculated for four different physicians for each patient and compared. The OARs do not have much variability across physicians, and the OAR contour drawn manually by the original treating physician is used for dose prediction for all cases.

## 2. MATERIALS AND METHODS

## 2.1. DATA

This study includes retrospectively collected CT volumes originally obtained during the initial simulation for adjuvant or salvage postoperative radiotherapy (PORT) to the prostatic fossa of 297 patients who had undergone resection for prostate cancer. These patients were treated with PORT by four expert genitourinary radiation oncologists at UT Southwestern Medical center between January 2010 and December 2017. Two hundred twenty patients had radiation delivered only to the prostate fossa (65-72 Gy dose), and 77 patients had radiation delivered to regional lymph nodes in addition to the prostate fossa (~41-45 Gy dose). These scans were contoured by four different physicians, and these contours were used for patient treatment. For each patient, the data available were the patient CT, masks of the planning target volumes and organs-at-risk, and the 3D dose distribution, which was generated from a VMAT setup. The PTV that was used for clinical dose planning and the resulting 3D dose distribution that was used for patient treatment are considered to be the ground truth PTV and dose for each patient throughout this paper. The structures available in the dataset were the PTVfossa, PTVnode, Body, Bladder, Rectum, Left Femoral Head, and Right Femoral Head. Each CT volume contains 60-360 slices and a voxel size of $1.17 \times 1.17 \times 2$ mm$^3$.

The dataset was split into 217 patients for training, 50 patients used for testing and 30 patients used for validation. All the 50 test patients had treatment only to prostatic fossa since for dosimetric impact analysis we are interested only in patients with a single prostatic fossa CTV without additional nodal targets.

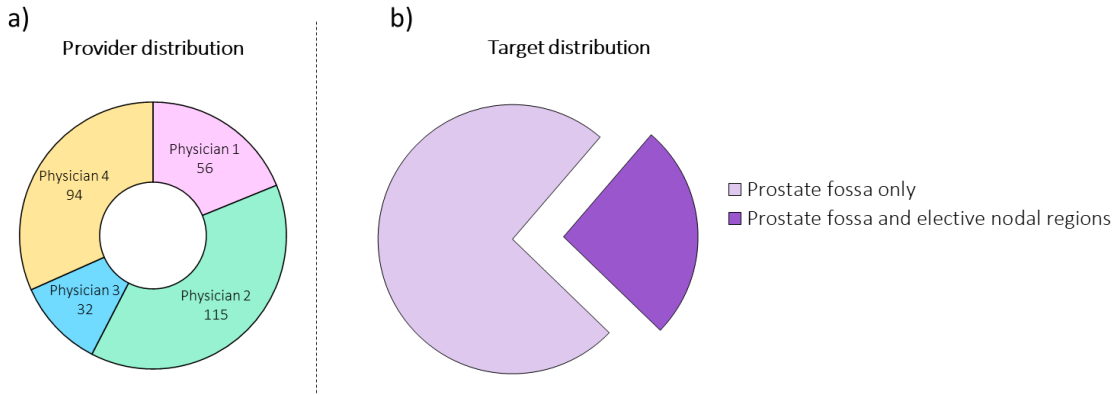

*Figure 1: Patient distribution information. a) Distribution of patients in the dataset across the four genitourinary physicians. b) Distribution of patients in the dataset across targets treated (PTVs). Patients with only prostate fossa as target have only one PTV, while patients with an additional nodal target have two PTVs.*

## 2.2. PTV SEGMENTATION NETWORK

Three different models were trained and compared for PTV segmentation. Model A is a volumetric U-Net [27] with residual blocks [28], Res-UNet, trained with only CT as inputs, and Model B is a Res-UNet trained with CTVs, OARs and CT as inputs. Model C is a physician-specific model, and is a set of four models, one for each physician. Each of these physician models is a Res-UNet with CTV, OARs, and CT as inputs, and is trained separately on each of the four physician's data.

To keep the study as fair as possible, all models trained utilized the same architecture (Figure 2) and hyperparameters. The model starts by performing a convolution with a kernel size of $3 \times 3 \times 3$, followed by ReLU, Group Norm (GN) [30], and then Dropout (DO) [29]. In this paper, we will refer to this set of operations as Conv-ReLU-GN-DO. This is followed by a residual block that incorporates ReLU activation and GN, then by a maxpooling operation ($2 \times 2 \times 2$). These sets of calculations are performed three times to reach the bottom of the U-Net. The bottom layer has two sets of atrous convolutions, which allow us to enlarge the field of view of filters to incorporate the larger context. Then, the features are upsampled, which consists of a $2 \times 2 \times 2$ upscale and two sets of Conv-ReLu-GN-DO operations. These upsampling operations are performed three times to return the data to its original resolution as the input. At each upsampling step, the features from the left side of the U-Net are copied over and concatenated with the upsampled features. Finally, the model uses a $1 \times 1 \times 1$ convolution to map the feature vector to the required number of classes and a connected sigmoid layer to output the probability value of the target. We used the Dice similarity coefficient loss as the loss function to train the network. Since the CTV contour was already available, CT volumes were cropped to $160 \times 160 \times 64$–sized volumes around the CTV before being input into the model. We used the Adam optimizer and switched to SGD in the last 30 epochs [31] with a learning rate of 0.001. All the models were trained on a V100 GPU with 32 GB of memory, using TensorFlow version 2.1. The batch size was set to 4 due to memory limitations. The ground truth PTV that was used for supervised training is the clinical PTV that was used for patient treatment.

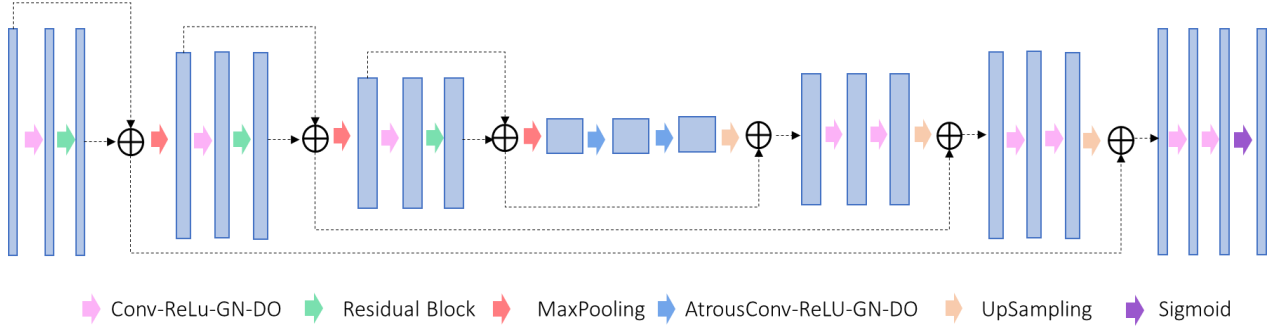

*Figure 2: Res-UNet used for PTV segmentation.*

## 2.3. DOSE PREDICTION NETWORK

For dose prediction, we utilized a U-Net–style architecture [27] (Figure 3). The model starts with Conv-ReLU-GN-DO, which is repeated twice. This is followed by a maxpooling operation ($2 \times 2 \times 2$). These sets of calculations are performed three times in total to reach the bottom of the U-Net. Then the features are upsampled via a $2 \times 2 \times 2$ upscale and two sets of Conv-ReLu-GN-DO operations. This set of operations is performed three times to return the data to its original resolution as the input. At each upsampling step, the features from the left side of the U-Net are copied over and concatenated with the upsampled features. The final layer has a linear activation and outputs the dose map. MSE is the most common loss function used for training regression models even though MAE is commonly used for evaluating the performance of models. The problem in using MAE loss (for neural nets especially) is that its gradient is the same throughout, which means the gradient will be large even for small loss values. This isn't good for learning. MSE behaves nicely in this case as the gradient of MSE loss is high for larger loss values and decreases as loss approaches 0, making it more precise. Huber loss approaches MSE when $\delta \sim 0$ and MAE when $\delta \sim \infty$ and is considered to be a good compromise between MSE and MAE. Huber loss is defined as,

$$H_\delta(x) = \begin{cases} \frac{1}{2}x^2 & for\ |x| < \delta \\ \delta\left(|x| - \frac{1}{2}\delta\right) & otherwise \end{cases} \quad (1)$$

From an optimization perspective, the Huber loss (Equation(1)) has similar characteristics to MAE, but is additionally differentiable at $x = 0$.

Since both MSE and Huber loss are viable loss functions, two models with the same architecture were trained: Model A with MSE as the loss function and Model B with Huber loss as the loss function. The final model, Model AB, is an ensemble of Model A and Model B. Both models were optimized via the Adam optimizer, which was switched for SGD in the last 30 epochs [31], with a learning rate of 0.001. The training was performed on a V100 GPU with 32 GB of memory, using TensorFlow version 2.1. The batch size was set to 1 due to memory limitations. The input to the model were PTV binary mask and signed distance arrays for OARs. Signed distance arrays for each OAR structure were generated by calculating Euclidean distance between the PTV boundary and each voxel on the OAR binary masks. Each voxel in the signed distance array represents the minimum distance from the closest edge of the PTV contour. OAR voxels outside the PTV contour had a positive distance value, and voxels inside the PTV had a negative distance value.

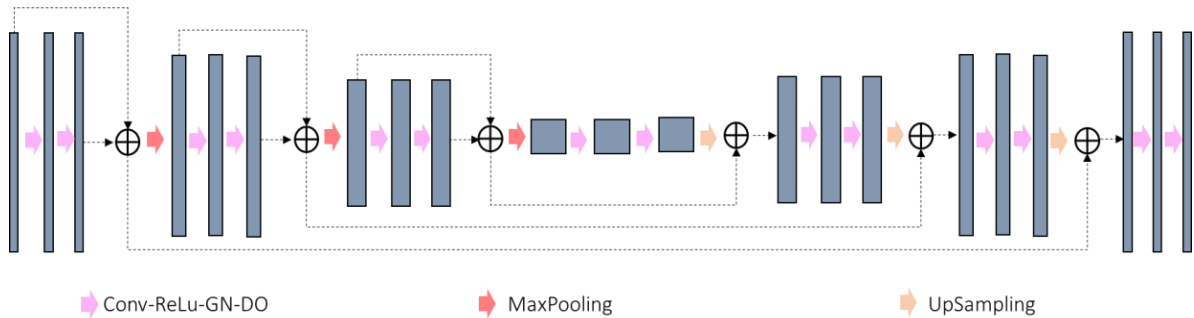

*Figure 3: Dose Prediction Network. Schematic of the U-Net used in the study.*

## 2.4. PERFORMANCE EVALUATION

### 2.4.1. PTV SEGMENTATION

For the test patients, we compared Dice similarity coefficients between each of the model predictions and the PTV that was used for treating the patient, which from here on will be referred to as clinical PTV. The recommended CTV–PTV expansion margin for postoperative prostate CTV at our institution is 9 mm anterior, 5 mm posterior, and 7 mm in all other directions. The clinical CTV was expanded per these margins and compared to the clinical PTV used to create a baseline.

### 2.4.2. DOSE PREDICTION

To evaluate the performance of the dose prediction models, we assessed their prediction accuracy on the 80 test patients, 50 with only prostate fossa targets and 30 with additional nodal targets. We evaluated the effects of the models on clinically relevant metrics, including the mean dose (Dmean) and max dose (Dmax) to each structure. Dmax is defined as the dose to 2% of the volume for each structure, as recommended by the ICRU-83 report [32]. We also evaluated the model performance of the frameworks on the total PTV dose coverage D99, D98, and D95, which are the doses to 99%, 98%, and 95%, respectively, of the total PTV volume. We assessed the PTV homogeneity ((D2-D98)/D50) for the total PTV.

### 2.4.3. PHYSICIAN DOSE COMPARISON

For evaluating the dosimetric impact of the physician CTV style variability, the in-house physician style-ware CTV segmentation model is used for predicting physician style CTVs for each patient. For each patient, we already have the CTV contour manually segmented by one physician that was used to create the ground truth dose plan and treat the patient, so we use the segmentation model to generate CTVs in the other three physicians' styles. Following this, the physician-specific PTV expansion DL network is used for expanding the three physician's CTVs to PTVs. The PTV expansion model is trained to emulate the way physicians expand the CTV to PTV. These PTVs are further fed into the dose prediction model to predict the dose distribution. To compare plans with PTVs from different physicians, we evaluated the OAR DVHs and compared clinically relevant dose metrics. We used the 50 test patients with only prostatic fossa CTVs for this comparison. For bladder and rectum, we measured Maximum Pt Dose (MaxPt), V70 and V60, which are the maximum dose to any point in the structure and the volume in cc of the structure that receives more than a 70 Gy dose and a 65 Gy dose, respectively. For left and right femoral heads, we measured MaxPt and V48, which is the volume in cc of the structure that receives more than a 48 Gy dose. We further calculate the overhead value for each physician. Overhead is the extra room left by the physicians for each dose metric below the corresponding limit as a % of the limit. The higher the overhead, the lower the value of the OAR dose metric and the lower the dose to the structure.

## 3. RESULTS

### 3.1. PTV SEGMENTATION

For PTV segmentation, we compared three models along with a recommended CTV–PTV margin expansion. Figure 4a shows a comparison of all the models in terms of Dice similarity coefficient. Model A, which does not use the CTV as an input, performed the worst, with an average DSC of $79.9 \pm 5.7\%$. Adding the CTV as an input to the model significantly improved the model accuracy. Model B has a DSC accuracy of $94.4 \pm 4.1\%$. The physician-specific models, Model C (four separate models), performed the best, with a DSC accuracy of $96 \pm 1.9\%$. Expanding the CTV with the recommended PTV margins gave an accuracy of $92.2 \pm 2.5\%$. Model C performed significantly better than the other DL models, as well as the recommended expansion (pairwise t-test, $p<.05$). Model C also had a much smaller standard deviation than the other models. The bar plot in Figure 4b shows how Model C outperformed the standard margin expansion for each of the 50 fossa only test patients. Since physicians do not follow standard margins due to toxicity concerns, an individual CTV-PTV expansion DL model for each physician works significantly better than a standard expansion. This makes the DL based PTV expansion model beneficial for studying dosimetric impact of physician style. Visual examples of the model performance also show that Model C (physician-specific model) performs better than Model B with better conformity to clinically used PTVs (Figure 4c). For physician style dosimetric impact evaluation that follows, the physician specific model is used for expanding from CTV to PTV for accurately emulating the workflow.

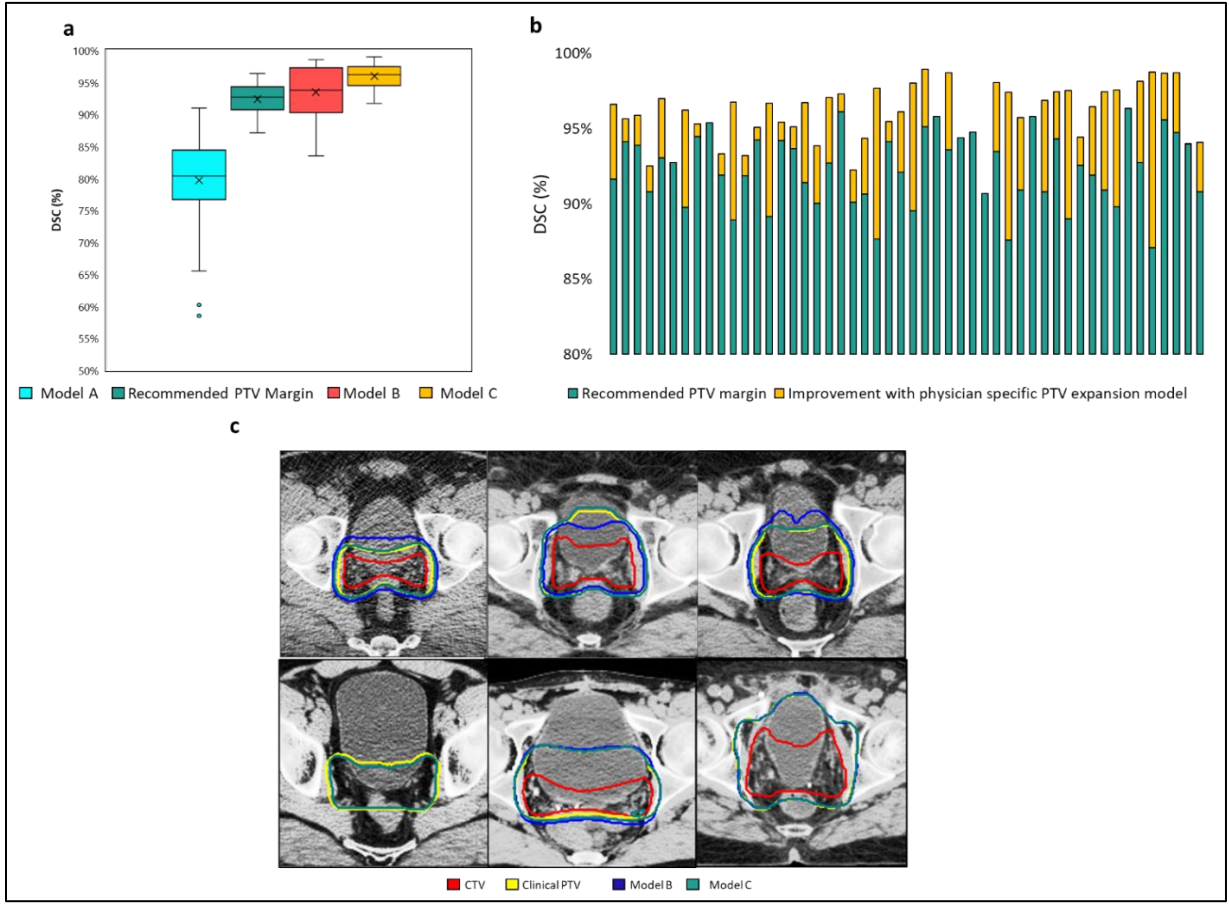

*Figure 4: PTV segmentation result comparison for Model A, which takes only CT as input; Model B, which takes CTV and OARs as input along with CT; and Model C, a superset of four models, each trained on a single physician's data, which uses CTV, OARs and CT as input. a) Box plots comparing DSC distribution for all the test patients for the considered models. b) Improvement in PTV accuracy with the physician-specific expansion model over recommended margin expansion for a subset of 50 test patients(only prostatic fossa CTV treated). c) Visual examples showing improvement with physician specific model over a combined data model.*

## 3.2. DOSE PREDICTION

We compared three models: Model A, which was trained with MSE loss; Model B, which was trained with Huber loss; and Model AB, which is an ensemble of Models A and B. Model B performed significantly better than Model A for the PTV (pairwise t-test, $p<.05$), in terms of the mean absolute error of Dmean, Dmax, D99, D98 and D95 (Figure 5a). Both models maintained a mean absolute error less than 5% of the prescription dose. By contrast, Model A performed significantly better than Model B for OARs (pair-wise t-test, $p<.05$), in terms of mean absolute error of Dmean and Dmax (Figure 5b). Both models maintained a mean absolute error less than 5% of the prescription dose for all OAR dose metrics.

Model AB, being the ensemble of Models A and B, performed significantly better (pairwise t-test, $p<.05$) than both Model A and Model B for both PTV and OARs. Model AB maintained a mean absolute error less than 3% of the prescription dose for all dose metrics. In terms of PTV homogeneity, with a perfect homogeneity being 0, the PTV dose predicted by Model A has a better conformity than the dose predicted by Model B, and the homogeneity of the ensemble model is similar to that of Model A (Figure 5c).

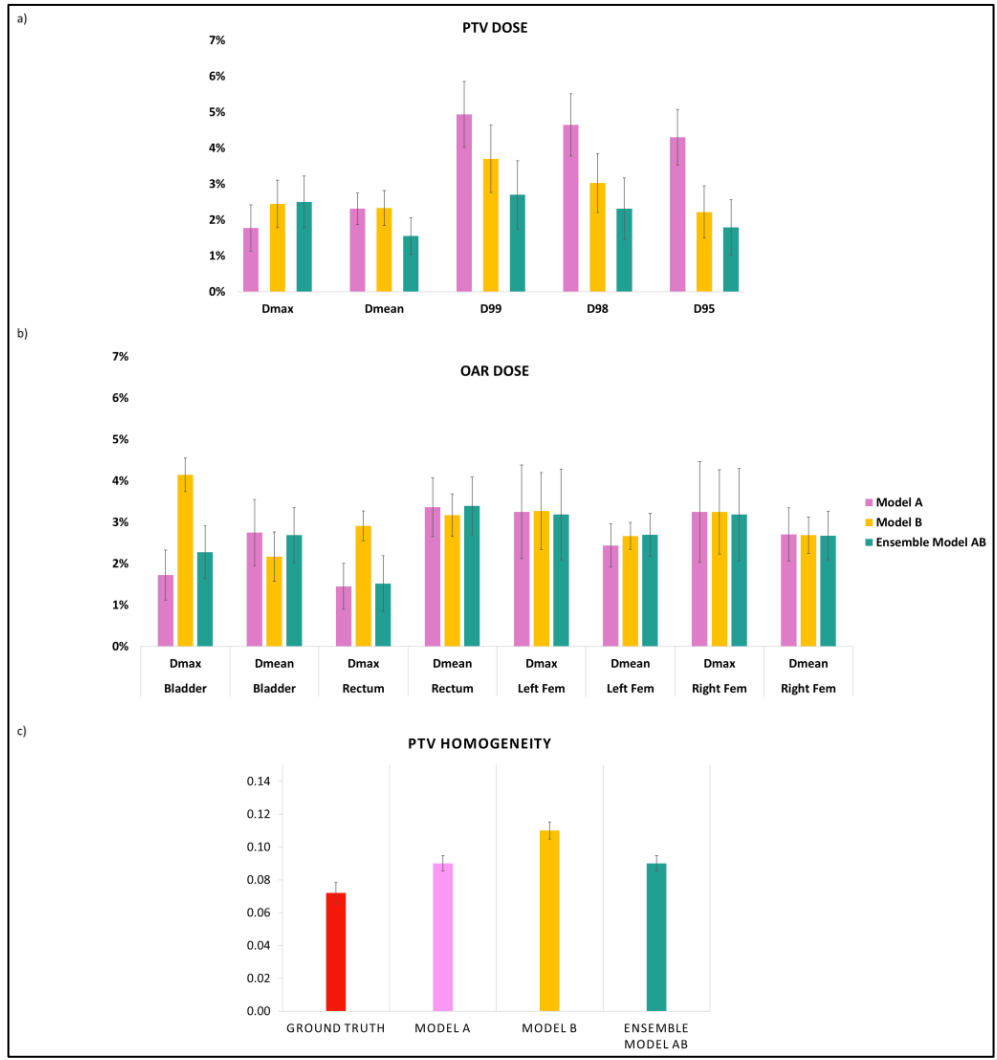

*Figure 5: Comparison of dose metrics for the three models. Model A is a U-Net trained with MSE loss, Model B is a U-Net trained with Huber loss, and Model AB is an ensemble of Models A&B. a) Mean absolute error of the dose coverage ($Dmax, Dmean, D_{99}, D_{98}, D_{95}$) between the predictions and the ground truth dose for PTV. b) Mean absolute error of the dose coverage ($Dmax$ and $Dmean$) between the predictions and the ground truth dose for OARs. c) Homogeneity of the Model A prediction, Model B prediction, Model AB prediction and ground truth. Error bar represents the 95% confidence interval $\left(\bar{x} \pm 1.96 \frac{\sigma}{\sqrt{n}}\right)$.*

Figure 6a shows an example DVH and dose wash for a dose plan for prostatic fossa only. In this case, there is only one PTV, and the model prediction is close to the ground truth dose. Figure 6b shows an example DVH and dose wash for a dose plan for both prostatic fossa and elective nodal regions. The predicted dose DVH as well as dose wash can be observed to be very close to the corresponding ground truth.

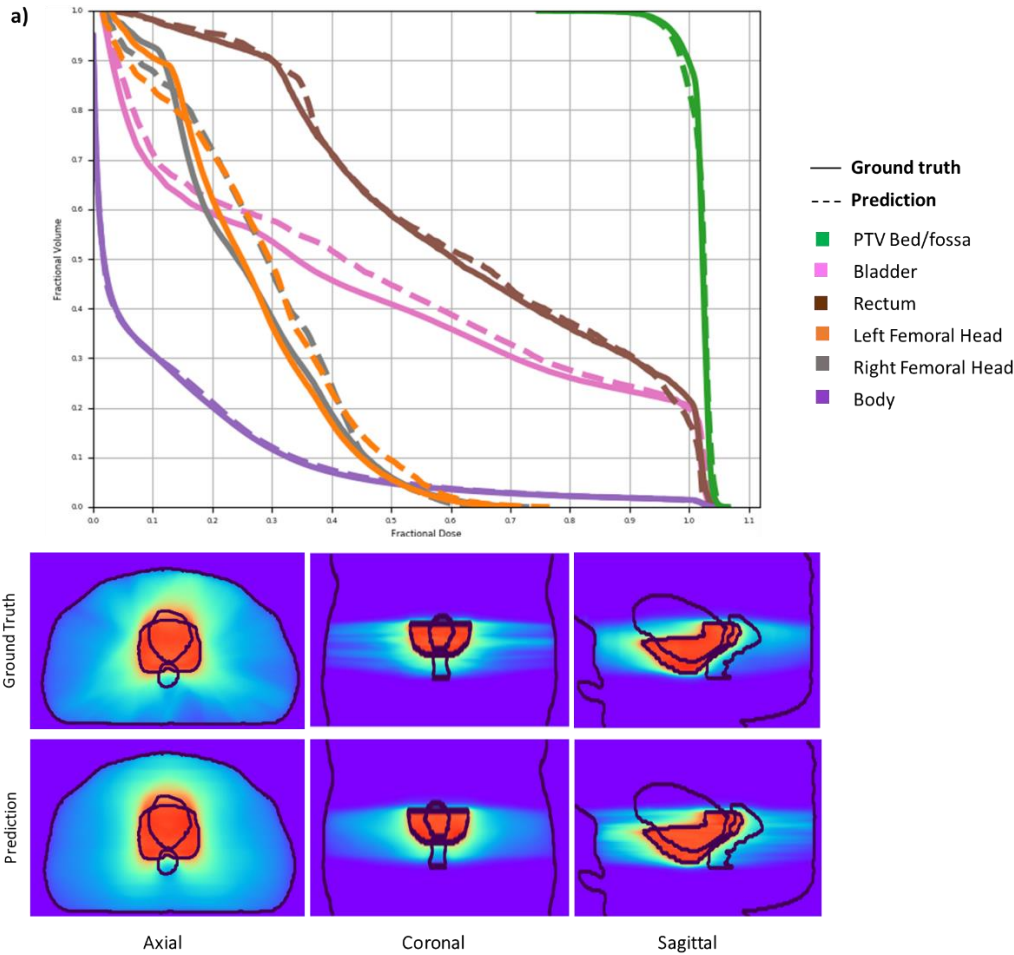

*Figure 6: DVHs (top row) and dose washes (bottom row) for an example test patient with the ground truth and the model prediction. The predicted dose DVH as well as dose wash can be observed to be very close to the corresponding ground truth.*

## 3.3. PHYSICIAN STYLE DOSE COMPARISON

We compared the dose to OARs for each test patient when using different physician CTVs. Model C from 3.1 which is the physician specific CTV-PTV expansion model was used for expanding the CTV for each physician to the corresponding PTVs. Model AB which is the ensemble model from 3.2 was used for calculating the dose. Figure 7 depicts the values of OAR volumetric dose metrics for all test patients. The dashed line shows the limit for the metrics; the value should be below this limit to avoid any toxicity to the tissues. We can see that the CTV variations affect the OAR dose. This is more prominent in some patients than in others. But whichever CTV is used, the volumetric dose constraints metrics seem to have a large overhead, which implies that the risk of significant toxicity is within acceptable limits for any physicians' CTV. On the other hand, the MaxPt [Gy] has large variations across physicians. For Rectum and Bladder, the MaxPt seems to be within limits for all patients except one. For the femoral heads, on the other hand, MaxPt dose has large fluctuations and seems to go over the the limit for 8 out of 50 patients for Left femoral head and 14 out of 50 patients for Right femoral head.

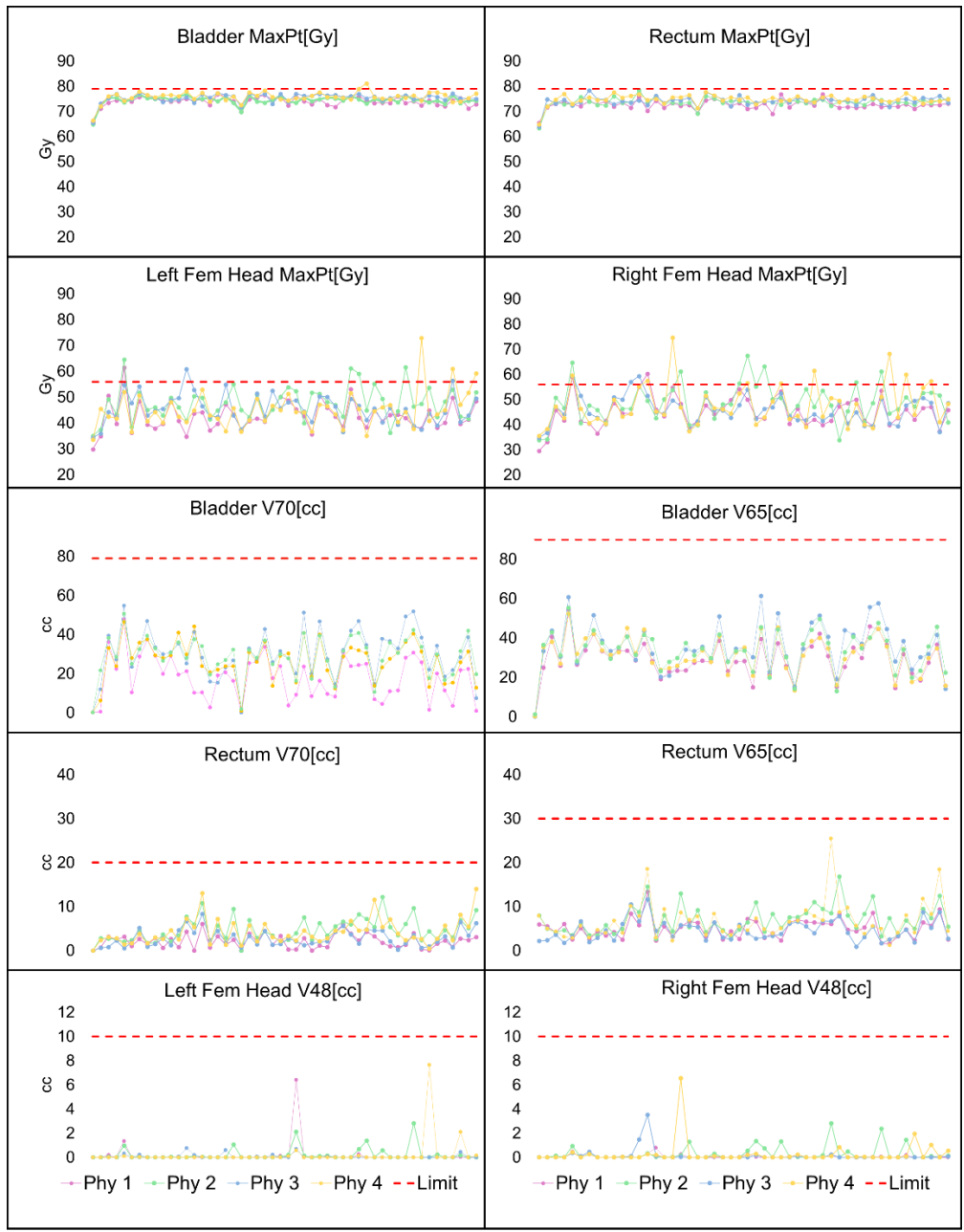

*Figure 7: Scatter plots depicting the values of OAR volumetric dose metrics for four different physician CTVs for all 50 test patients.*

Figure 8 shows dose variation for example patient with different physician CTVs. Figure 8a shows a bar plot with dose and volume constraints values when dose prediction is performed using each physicians' CTV. Even with the stylistic variations in contouring across physicians, the doses to OARs can be observed to be still within the recommended limit. Figure 8b shows the corresponding DVHs for bladder, rectum, and femoral heads. The DVHs slightly vary especially for rectum but the values still fall within the recommended limits. Figure 8c shows an example of the four dose maps along with their dose metrics for the test patient.

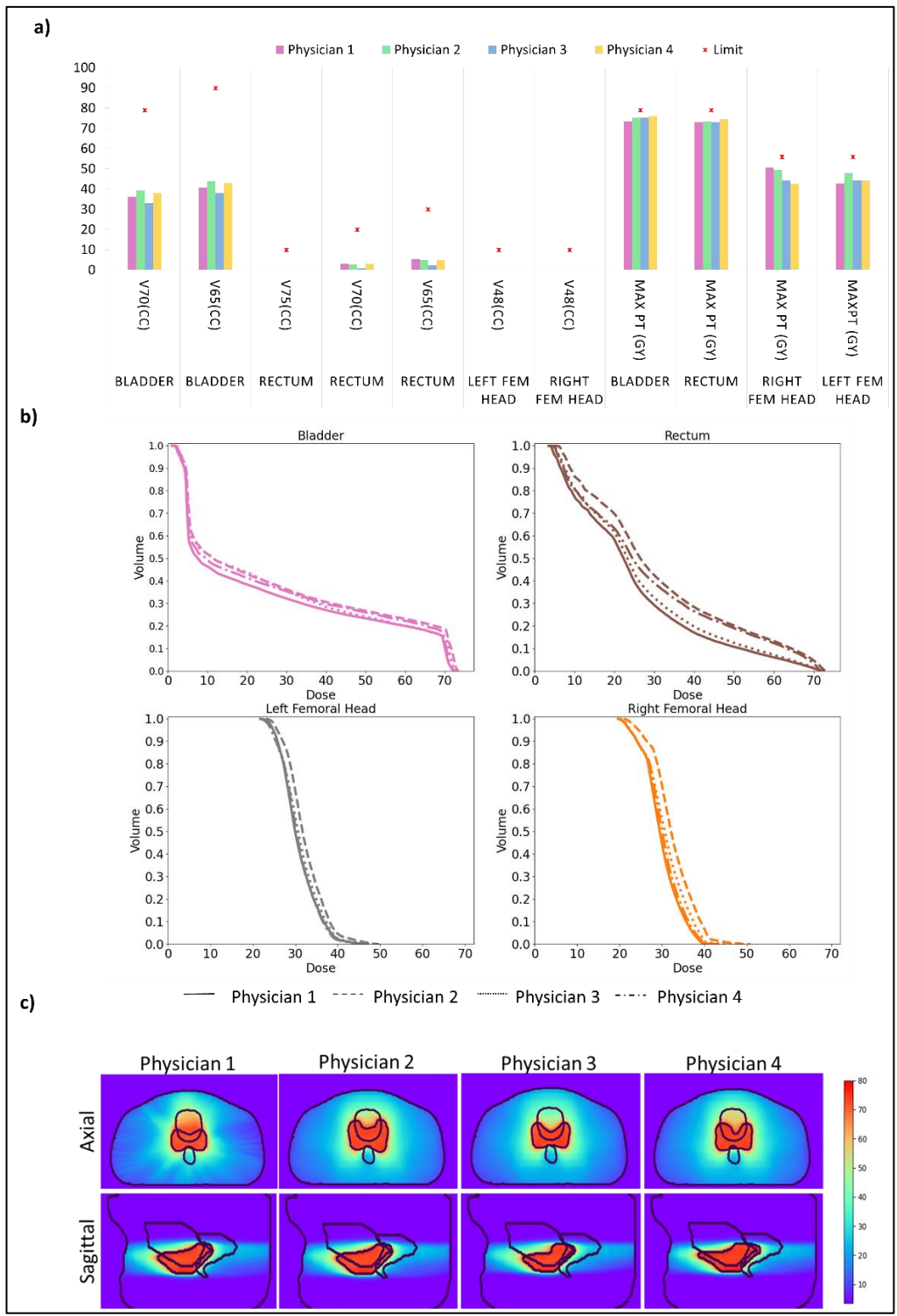

*Figure 8: For an example patient: a) dose and volume metrics values when dose prediction is performed using each physicians' CTV. b) predicted DVHs for bladder, rectum, and femoral heads when planning is performed with different physician CTVs Vs. c) The dose washes (the same axial and sagittal locations) show the variation in dose with PTV, Bladder and Rectum contours in black.*

Figure 9 shows the average overhead for each physician for each OAR. Overhead is the extra room that physicians left for a volumetric dose constraint below the corresponding limit as a percentage of the limit. The higher the overhead, the lower the value of the OAR volumetric dose constraint and the lower the dose to the structure. On an average, physician 1 was observed to have higher overhead for Bladder showing lower dose to Bladder with physician 1's CTV contouring and PTV expansion style. p-value calculated using paired t-test for means shows that the dose reduction with physician 1's contour is significant. Physicians 2 was observed to have the lowest overhead for Bladder, suggesting higher dose to OARs for physician 2's CTV contouring and PTV expansion style. p-value calculated using paired t-test for means shows that the dose increase with physician 2's contour is significant. For Rectum, physicians 1 and 3 were observed to have significantly higher overhead compared to Physicians 2 and 4. For Left femoral head, all physicians were observed to have statistically indifferent overhead average. For Right femoral head, physician 1 was observed to have statistically significant higher overhead compared to physician 2. Combining for all the OARs, physician 1's CTV contouring and PTV expansion style gives the highest overhead values and hence lowest dose to OARs. This stems from the difference in the way physicians deal with trade-off between high dose to pathologically adverse areas and OAR dose. Lower OAR dose is better but as long the dose constraints are met, the risk of significant toxicity is within acceptable limits for any physician.

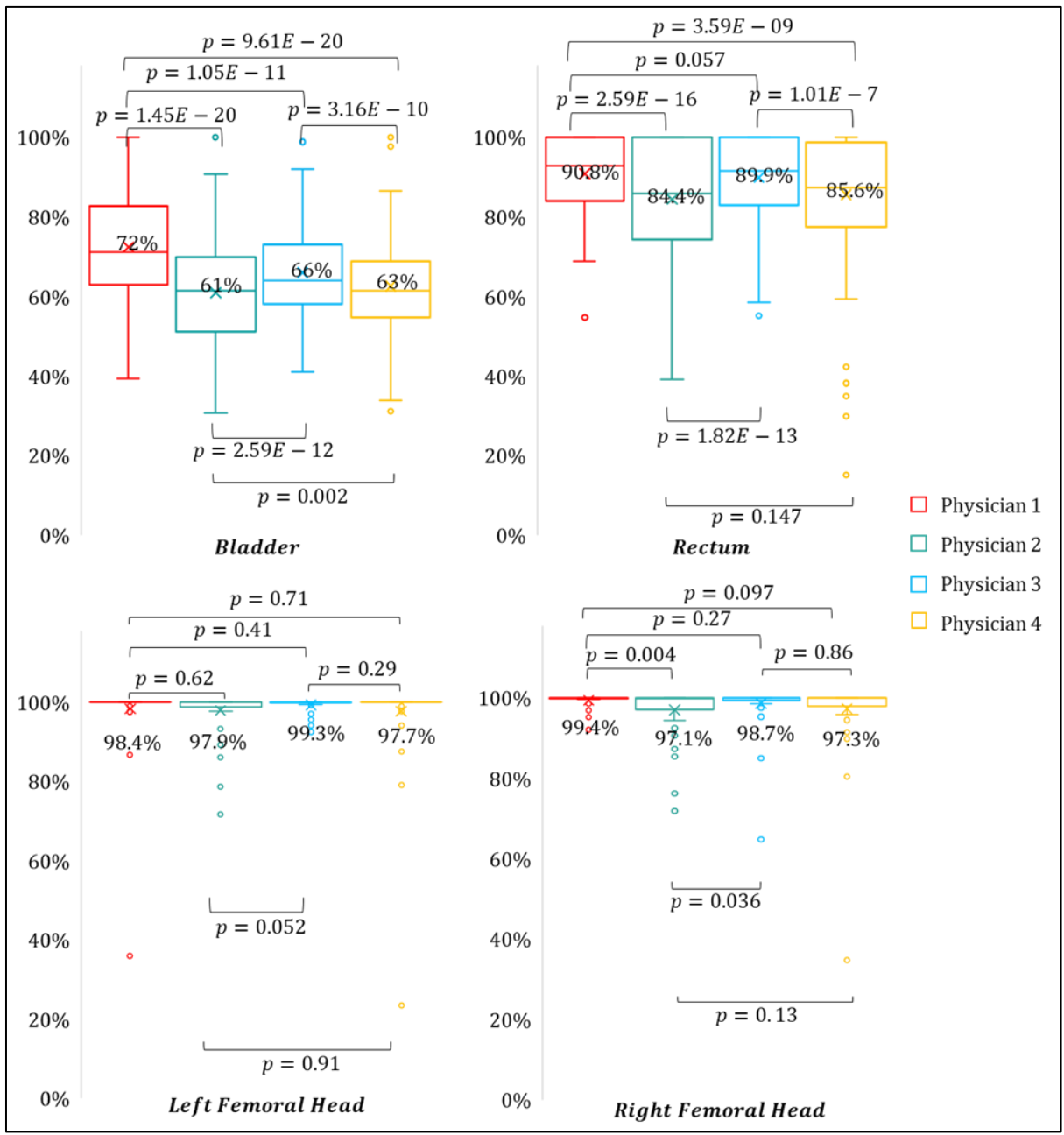

*Figure 9: overhead for each physician for each OAR (metrics combined). Overhead is the extra room left by the physicians for each dose metric below the corresponding limit as a % of the limit. The higher the overhead, the lower the value of the OAR dose metric and the lower the dose to the structure. p-values determined using paired t-test for means between each physicians is noted on the figures. p<0.05 shows significance. Mean value is printed next to the corresponding boxes."*.

## 4. DISCUSSION

In postoperative prostate patients, the CTV is not an anatomically established structure but rather one determined by the physician based on the clinical guideline used, the preferred tradeoff between tumor control and toxicity, their experience and training background, and other factors. This results in high inter-observer variability between physicians. Inter-observer variability has been considered an issue, but its dosimetric consequences are still unclear, because the significant amount of time required for dose planning makes it impractical to produce multiple physician CTV contours for each patient. In this study, we leverage deep learning to simulate the clinical workflow for postoperative prostate treatment planning to evaluate the dosimetric impact of physician style variations on organs-at-risk.

In this work, we have designed a dose prediction tool that can predict dose with less than 3% error. We compared three models: a model trained with MSE loss, a model trained with Huber loss and an ensemble of the two models. Although the model trained with MSE resulted in better PTV dose metrics, the model trained with Huber loss resulted in better OAR dose metrics. The ensemble model performed better than or similar to both individual models. To our knowledge, this is the first DL-based radiotherapy dose prediction study for the postoperative setting.

To demonstrate the clinical use case of this dose prediction model, we leveraged an in-house physician-specific CTV segmentation model that can create physician style–aware segmentations. Additionally, since PTVs are required for dose planning, and there is no standard CTV–PTV margin used by all physicians, we designed a PTV segmentation model that can produce accurate PTVs from the CTV. We designed three models and compared them for PTV segmentation. All three models used an encoder-decoder architecture with group normalizations, atrous convolutions, residual blocks and skip connections between the encoder and the decoder. The model that was trained separately on each physician's data, with CT, CTV, and OARs as input, performed the best, with 96% DSC accuracy. This model performed significantly better than using a standard recommended margin of 9 mm anterior expansion, 5 mm posterior expansion, and 7 mm expansion in all other directions. When comparing the OAR dose metrics, we observed that even though different physician PTVs have a significant impact on the OAR dose, they all fall below the limits for the volume constraints. The MaxPt dose being a single point value could have large stray effects from errors in the dose prediction model, and the effect of this error on MaxPt dose needs to be studied further. The MaxPt stray effect was observed to be higher for femoral heads as these compared to Bladder and Rectum are farther away from the target.

Physician 1 has the highest metric overhead values while Physician has the lowest metric overhead values suggesting that Physician 1's CTV contouring and PTV expansion style gives the lowest dose to OARs while physician 2's CTV contouring and PTV expansion style gives the highest dose to OARs but all the metrics were observed to be within limits for most of the patients. CTV contouring by physicians is always a trade-off between dose to pathologically relevant areas vs dose to OARs. Physicians rely on their experience and knowledge to make that decision. Even though physician 1's CTV contour results in lowest OAR dose, for all the physician's OAR doses fall below the limits for the volume constraints.

This study has limitations, most of which come from the simulation errors. There are three deep learning models in play, so the sequential errors would add up. Nevertheless, these errors are still acceptable, considering that this is a simulation study. Performing the segmentation, the margin expansion or the dose planning manually could make the comparison more accurate. In future studies, including deep learning model uncertainties would help in evaluating the error associated with each prediction. The results and corresponding conclusions are only recommendations based on a simulation study. A dosimetric impact study would be difficult to conduct without simulations owing to the significant amount of time it would take to dose plan for enough patients that could give a statistical conclusion.

# 5. CONCLUSION

In this study, we have simulated the clinical workflow for postoperative prostate treatment planning and evaluated the dosimetric impact of physician style variations on organs-at-risk. We have developed and proposed a model for predicting volumetric dose for postoperative prostate cancer patients. Using our proposed implementation, we can accurately predict the dose distribution from the PTV and OAR contours, and the prescription dose. On average, our proposed model predicted the OAR max dose within 4% and the mean dose within 3% of the prescription dose on the test data. We have also developed an accurate CTV–PTV expansion model that has 96% DSC accuracy. Physician 1 was observed to leave the largest room for OAR dose metrics suggesting that Physician 1's CTV contouring and PTV expansion style gives the lowest dose to OARs, but all the metrics were observed to be within limits for most of the patients.

## DATA AVAILABILITY

All the datasets were collected from one institution and are non-public. In accordance with HIPAA policy, access to the datasets will be granted on a case by case basis upon submission of a request to the corresponding authors and the institution.

## CODE AVAILABILITY

The DL models will be free to download for non-commercial research purposes on GitHub after paper acceptance. (https://github.com/anjali91-DL/Post-op-prostate-doseprediction-model)

## AUTHOR CONTRIBUTIONS

All authors have made contributions to the manuscript, including its conception and design, the analysis of the data and the writing of the manuscript. All authors have reviewed all parts of the manuscript and take responsibility for its content and approve its publication.

## CONFLICT OF INTEREST

The authors declare no competing financial interest. The authors confirm that all funding sources supporting the work and all institutions or people who contributed to the work, but who do not meet the criteria for authorship, are acknowledged. The authors also confirm that all commercial affiliations, stock ownership, equity interests or patent licensing arrangements that could be considered to pose a financial conflict of interest in connection with the work have been disclosed.

## ACKNOWLEDGMENTS

We would like to thank Dr. Jonathan Feinberg for editing the manuscript and Varian Medical Systems, Inc. for providing funding support.